\documentclass[aps,prb,preprint,groupedaddress,showpacs,showkeys]{revtex4}
\usepackage[dvips]{graphicx}
\usepackage{amsmath}
\begin{document}
\title{Incommensurability and edge states \\
in the one-dimensional $S=1$ bilinear-biquadratic model}
\author{Takahiro Murashima and Kiyohide Nomura}
\affiliation{Department of Physics, Kyushu University, 
Fukuoka 812-8581,Japan}
\date{\today}
\begin{abstract}
Commensurate-incommensurate change
on
the one-dimensional $S=1$ bilinear-biquadratic model
(${\cal H}(\alpha)=\sum_i \{
{\bf S}_i\cdot {\bf S}_{i+1} +\alpha ({\bf
 S}_i\cdot{\bf S}_{i+1})^2\}$)
is examined.
The gapped Haldane phase 
has two subphases (the commensurate Haldane subphase and 
the incommensurate Haldane subphase)
and the commensurate-incommensurate change point
(the Affleck-Kennedy-Lieb-Tasaki point, $\alpha=1/3$).
There have been two different analytical predictions about
the static structure factor
in the neighborhood of this point.
By using the S{\o}rensen-Affleck prescription,
these static structure factors
are related to the Green functions,
and also to the energy gap behaviors.
Numerical calculations support one of the predictions.
Accordingly,
the commensurate-incommensurate change is
recognized as a motion of a pair of poles in the complex plane.
\end{abstract}

\maketitle

\section{Introduction\label{sec_Intro}}
Commensurate-incommensurate (C-IC) transitions 
induced by frustration
are important problems in many-body quantum spin systems.
Among them,
a C-IC change with an excitation gap is observed in one-dimensional
(1D) quantum spin models.
\cite{Tonegawa-Harada,
Tonegawa-Kaburagi-Ichikawa-Harada,Bursill-Xiang-Gehring}
This change is not a phase transition without an excitation gap.
Whereas theories of C-IC transitions with no excitation gap
({\it e.g.} the Pokrovsky-Talapov transition\cite{Pokrovsky-Talapov})
have been developed,
those of C-IC changes for quantum systems have not been yet.
In some classical systems, 
analytical approaches to the C-IC change
have been discussed,\cite{Stephenson1970a, Stephenson1970b}
and then
a random phase approximation approach
has been succeeded phenomenologically.\cite{GM1986}
However, on the one hand the 1D frustrated Ising model for finite
temperature
cannot be mapped onto the 1D quantum case,
on the other hand the transfer matrix for the 2D Ising model on the
triangular lattice is non-symmetric, thus its correspondence to the 1D
quantum case is not a simple problem. Therefore, an independent
analytical research for the 1D quantum C-IC change is needed.

There are typical quantum models
which show the C-IC change;
the 1D $S=1/2$ next-nearest-neighbor (NNN) model\cite{Tonegawa-Harada}
and the 1D $S=1$ bilinear-biquadratic (BLBQ) model.\cite{Bursill-Xiang-Gehring}
It is common between these models that
the C-IC change occurs at the solvable point;
the Majumdar-Ghosh point\cite{M-G}
in the 1D $S=1/2$ NNN model
and the Affleck-Kennedy-Lieb-Tasaki (AKLT) point\cite{AKLT1,AKLT2}
in the 1D $S=1$ BLBQ model.
These solvable points are called 
as the disordered point.\cite{Stephenson1970a, Stephenson1970b}
At the disordered point,
the correlation length is the smallest
and the ground state is described by the matrix product
state.\cite{M-G,AKLT1,AKLT2}
The correlation length and the incommensurate wave number
are not differentiable at the disordered point,
although they are continuous.
The structure factor (the Fourier transform of the
correlation function)
varies from the 2D Ornstein-Zernicke type 
(the modified Bessel function) in
the commensurate and incommensurate regions
to the 1D Ornstein-Zernicke type 
(the pure exponential function) at the
disordered point.\cite{Schollwock-Jolicoeur-Garel}

Recently,
in order to explain the C-IC change,
some analytical studies have been proposed.
F{\'a}th and S{\"u}t{\H o} have suggested that the C-IC change
occurs because of the existence of higher derivatives 
in an effective Lagrangian of the 1D $S=1$ BLBQ model.\cite{Fath-Suto}
On the other hand,
one of us (KN) has discussed 
the static structure factor.\cite{Nomura2003}
These studies
show two candidates for
the static structure factor,
although
they do not necessarily decide between them.

By the way,
S{\o}rensen and Affleck (S-A) have studied
two spin correlations and
energy gaps between the triplet and singlet states  
under the open boundary condition
by means of field theoretic approaches,\cite{Sorensen-Affleck}
although they have not concerned with the C-IC change.
Applying the S-A method to the C-IC problem,
we can calculate some parameters included 
in the dynamical structure factor,
{\it i.e.} the Green function.
In our previous paper,\cite{Nomura-Murashima}
we have already found 
that the incommensurate wave number can be calculated by 
the energy gap of edge states.
In this paper,
we attempt to determine the Green function.
After that,
the relation between
the singularities in the Green function and the incommensurability will
be clear,
and then
we will obtain a unified view among commensurate and incommensurate behaviors.

\begin{figure}
\includegraphics[width=8cm]{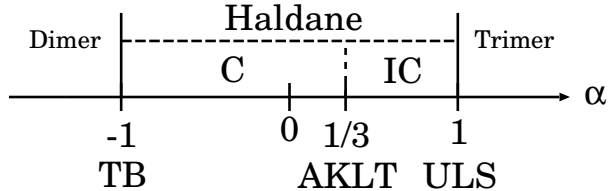}
\caption{Ground state phase diagram of the $S=1$ 
bilinear-biquadratic model.}
\label{fig_phase}
\end{figure}

In this stage, we summarize some known properties of
the 1D $S=1$ BLBQ model with the Hamiltonian;
\begin{equation}
{\cal H}(\alpha)=\sum_{i}\{ {\bf S}_i\cdot{\bf S}_{i+1} +
\alpha ({\bf S}_i\cdot{\bf S}_{i+1})^2 \}. \label{H1}
\end{equation}
The ground state phase diagram of this model is shown in Fig. \ref{fig_phase}.
This model is 
solvable at the AKLT point
\cite{AKLT1,AKLT2}
$\alpha=\alpha_{\rm D}=1/3$.
The ground state is the Valence-Bond-Solid (VBS) state
with the lowest excitation gap at the mode $k=\pi$.
One calls a phase,
the ground state of which is a unique disordered ground state
with a finite gap to the excited states,
as the Haldane phase after Haldane's conjecture.\cite{Haldane}
This phase extends from the Takhtajan-Bubujian (TB) point\cite{T,B1,B2}
$\alpha=-1$
to the Uimin-Lai-Sutherland (ULS) point\cite{U,L,S}
$\alpha=1$. 
At the TB or ULS points,
the BLBQ model is also solvable,
and has the gapless ground state
with the soft mode $k=0,\pi$ or $k=0,\pm 2\pi/3$,\cite{Fath-Solyom}
respectively.
For $\alpha < -1$,
there is the gapped dimerized (Dimer) phase,\cite{Oitmaa,AKLT1,Solyom}
whereas the gapless trimerized (Trimer) phase for $\alpha > 1$.\cite{Solyom}
Between the AKLT point and the TB point,
the lowest excitation has the wave number $k=\pi$,
while
the lowest excitations have
the incommensurate wave number
$k_{\rm IC}$, $2\pi/3 < |k_{\rm IC}| \le \pi$ 
between the AKLT point and the ULS point.\cite{Golinelli}
The wave numbers of the lowest excitations are different
in these two regions,
since the C-IC change occurs at the AKLT point.
\cite{Bursill-Xiang-Gehring,Schollwock-Jolicoeur-Garel}
The Haldane phase, therefore, has two subphases;
the commensurate Haldane subphase for $-1<\alpha<1/3$ and
the incommensurate Haldane subphase for $1/3<\alpha<1$.

In addition,
the VBS state
becomes increasingly significant in connection with quantum
entanglements.
\cite{Verstraete-Martin-Cirac, Fan-Korepin, Verstraete-Cirac}
The entanglements have a close relation to the matrix product state
as well as
the C-IC change. 
Therefore,
it will be useful for an understanding of the quantum entanglements to 
investigate
near the AKLT point, {\it i.e.} the C-IC change point.

The organization of this paper is as follows.
In the next section,
we summarize essential points of
the static structure factor concerning the C-IC change.
The analyticity of the static structure factor
explains that the change 
between branch points and a pole in the static structure factor
corresponds to the C-IC change.
In Sec. \ref{sec_EG}
we discuss the relation between the edge states and the Green
function on the basis of the S-A prescription.
From the analysis of this section and Sec. \ref{sec_SSF},
we expect some behaviors of the energy gap of edge states.
Before we study the energy gap of edge states numerically,
we discuss the lattice effect
in Sec. \ref{sec_LI}.
In Sec. \ref{sec_NA},
we confirm the gap behavior of edge states numerically,
which is related to the Kennedy degeneracy.\cite{Kennedy}
The last section gives a summary and a discussion.

\section{Static structure factor and incommensurability \label{sec_SSF}}
In our previous papers,\cite{Nomura2003,Nomura-Murashima}
we have discussed the functional forms of the static structure factor
concerning 
the C-IC change.
Before studying the relation between edge states and the C-IC change,
let us briefly summarize the essential points of
the static structure factor about the C-IC change.

\subsection{Analyticity of the static structure factor\label{subsec_ssf}}
From previous numerical results,
especially in Ref. [\onlinecite{Schollwock-Jolicoeur-Garel}],
one can find the static structure factor in each region as follows:
\begin{itemize}
\item[1.] In the commensurate region ($\alpha \ll \alpha_{\rm D}$),
\begin{equation}
S(q)\propto \frac{1}{\sqrt{q^2+m^2}}. \label{eq_ssf_com}
\end{equation}
\item[2.] At the disordered point ($\alpha=\alpha_{\rm D}$),
\begin{equation}
S(q)\propto \frac{1}{q^2 + m^2}. \label{eq_ssf_aklt}
\end{equation}
\item[3.] In the incommensurate region ($\alpha \gg \alpha_{\rm D}$),
\begin{equation}
S(q)\propto \frac{1}{\sqrt{(q-q_{\rm IC})^2 + m^2}} 
+ \frac{1}{\sqrt{(q+q_{\rm IC})^2 + m^2}}. \label{eq_ssf_incom}
\end{equation}
\end{itemize}
However,
one cannot connect these three expressions continuously.

Considering an analytic continuation of real $S(q)$ to the complex
plane,
we can discuss $S(q)$ in the complex $q$ plane.
In terms of the singularity in the complex $q$ plane,
there are poles at the disordered point,
in contrast to branch cuts in the other regions.

\begin{figure}
\includegraphics[width=8cm]{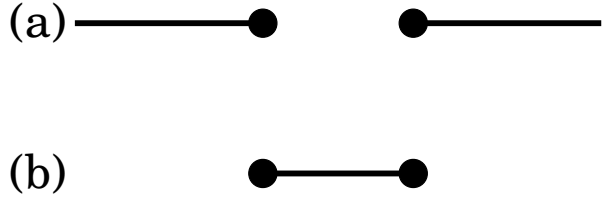}
\caption{Typical branch cuts of $f(z)=(z^2-d)^{-1/2}$. (a) $f(-z)=f(z)$.
(b) $f(-z)=-f(z)$.}
\label{fig_branch2}
\end{figure}

In order to unify these three expressions,
we reconsider the relation between a pole and a branch cut.
Considering the next function,
we can transform a pole into a branch cut, and vice versa,
\begin{equation}
f(z)\equiv (z^2 - d)^{-1/2}, \label{eq_fz}
\end{equation}
where $d$ is a real parameter.
This function has two branch points.
Typical branch cuts of $f(z)$ are shown in Fig. \ref{fig_branch2}.
In the case of the branch cuts (a), which connect each of branch points to
infinite distance,
$f(z)$ can be expanded in a Laurent series near $z=0$,
and then $f(z)$ is found to be an even function $f(-z)=f(z)$.
On the other hand,
in the case of the branch cut (b)
which connects both of branch points,
$f(z)$ is an odd function $f(-z)=-f(z)$
since $f(z)$ can be expanded at infinite distance 
[see Appendix \ref{app_dvf} in detail].
When $d=0$,
a simple pole appears in the case (b),
whereas the branch cuts remain in the case (a).
Thus we select the branch cut (b), 
and then deal with $f(z)$ as an odd function.
Then we find
$f(q-\widetilde{m}i)$ satisfies
\begin{align} 
\overline{f(\overline{q}+\widetilde{m}i)} &= f(q - \widetilde{m}i),\\
f((-q)+\widetilde{m}i) &= -f(q-\widetilde{m}i), 
\end{align}
where $\widetilde{m}$ is a real parameter.
Note that $q$, $\overline{q}$, and $-q$ belong to the same Riemann sheet.

The static structure factor must
satisfy several physical requirements (PRI):
\begin{itemize}
\item[1.] [reality on the real axis] 
$S(q)=\overline{S(\overline{q})}$.
\item[2.] [parity]
$S(q)=S(-q)$.
\item[3.] [algebraic singularity]
$S(q)$ is an analytic function of a complex variable $q$
except for several algebraic singular points.
\item[4.] [analyticity on the real axis]
Singular points and branch cuts must not cross the real
axis.
\end{itemize}
The above requirements represent properties of $S(q)$
on a fixed $\alpha$.
In addition,
\begin{itemize}
\item[5.] [$\alpha$-dependency near $\alpha_{\rm D}$]
$S(q)$ is an analytic function of a real parameter $\alpha$
in the neighborhood of $\alpha_{\rm D}$.
\item[6.] [property at $\alpha_{\rm D}$]
$S(q)$ is described with two simple poles
at the disordered point.

\end{itemize}

\begin{figure}
\includegraphics[width=8cm]{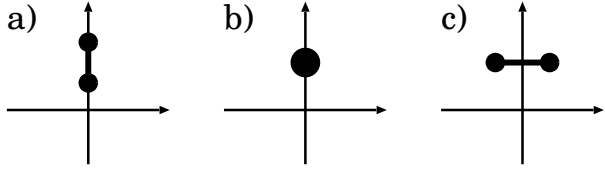}
\caption{Singularities of $f(q-\widetilde{m}i)$ when a) $d<0$, b) $d=0$, and
 c) $d>0$.}
\label{fig_fqim}
\end{figure}

On the basis of these requirements,
we can obtain two possible candidates of 
the static structure factor near the disordered point:
\begin{align}
S_{\rm sing}(q)&=
 Af(q+\widetilde{m}i)f(q-\widetilde{m}i),\label{eq_prod} \\
\intertext{or}
S_{\rm sing}(q)
&=A\frac{i}{2\widetilde{m}}[f(q+\widetilde{m}i)-f(q-\widetilde{m}i)], \label{eq_diff}
\end{align}
where real parameters $A$,
$\widetilde{m}$,
and
$d$ depend on $\alpha$.\cite{note-sum}
Figure \ref{fig_fqim} shows singularities of $f(q-\widetilde{m}i)$ when
a) $d<0$,
b) $d=0$,
and c) $d>0$.
$\widetilde{m}$ represents a distance between the real axis and the center of two
branch points.
Equations \eqref{eq_prod} and \eqref{eq_diff}
tend to $1/q^2$ at $q$ $\to$ $\infty$ limit.
The pre-factor $Ai/2\widetilde{m}$ in the difference type function
\eqref{eq_diff}
is determined
so that $S_{\rm sing}(q)=A/(q^2 + \widetilde{m}^2)$ when $d=0$.
Equation (\ref{eq_prod}) is the same one
which has firstly been proposed by
F{\'a}th and S{\"u}t{\H o} \cite{Fath-Suto} and
the other (\ref{eq_diff}) is discussed by KN.\cite{Nomura2003}
We would like to clarify the behavior of 
the static structure factor $S(q)$ by using another approach.
In the following sections,
we will investigate which is more appropriate structure factor 
either Eq. \eqref{eq_prod} or \eqref{eq_diff}.

\subsection{$\alpha$ dependency\label{subsec_taylor}}
In addition,
we can discuss how parameters $\widetilde{m},d$,
and $A$ 
depend on $\alpha$.
Since the correlation decays purely exponentially at the disordered
point,
we obtain $d=0, \widetilde{m}> 0$ at $\alpha=\alpha_{\rm D}$.
Generally, 
the requirement for the amplitude is $A\ne 0$
since the correlation function becomes perfectly zero for $A=0$.
Near $\alpha_{\rm D}$ we then expect that
$d$, $\widetilde{m}$, and $A$
can be expanded in a Taylor series:
\begin{align}
d&=d_1(\alpha-\alpha_{\rm D})+d_2(\alpha-\alpha_{\rm D})^2
+O((\alpha-\alpha_{\rm D})^3), \label{eq_d}\\
\widetilde{m}&=\widetilde{m}_0 + \widetilde{m}_1(\alpha-\alpha_{\rm D})
+O((\alpha-\alpha_{\rm D})^2), \label{eq_m}\\
\intertext{and}
A&=A_0 + A_1 (\alpha-\alpha_{\rm D}) +O((\alpha-\alpha_{\rm D})^2).
\end{align}
Besides,
PRI-4 in Sec. \ref{subsec_ssf}
means that $\widetilde{m}> \sqrt{-d}$ when $d<0$.

The incommensurate wavenumber $q_{\rm IC}\equiv \sqrt{d}$, therefore,
 behaves as
\begin{equation}
q_{\rm IC}
=
\sqrt{\alpha - \alpha_{\rm D}}
\sqrt{d_1 + d_2(\alpha - \alpha_{\rm D})},
\end{equation}
in the incommensurate region,
and $q_{\rm IC}=0$ in the commensurate region.
On the other hand,
the correlation length $\xi$,
which is related to the closest singular point to the real axis,
behaves as
\begin{equation}
\xi^{-1} \propto \widetilde{m} - \sqrt{-d},
\end{equation}
in the commensurate region,
and
\begin{equation}
\xi^{-1} \propto \widetilde{m},
\end{equation}
in the incommensurate region.

\subsection{Numerical difficulties in dealing with the static structure factor}
The previous consideration results that
the static structure factor should be Eq. \eqref{eq_prod} or
\eqref{eq_diff}.
To select one from two possibilities,
we may calculate numerically the correlation function with the DMRG method.
However,
there are some difficulties in dealing with the static structure factor
directly.
We require
\begin{itemize}
\item[1)] to calculate
a long range correlation near the disordered point
since the incommensurate wave number is small,
although the correlation length is short,
\item[2)] to consider how to avoid edge effects,
\end{itemize}
and also, 
\begin{itemize}
\item[3)] to improve accuracy in calculating 
the correlation function,
since the correlation function is less accurate 
than the energy eigen values.
\end{itemize}

Though it is indirect,
there is another approach which
uses the energy eigen values under the open boundary condition (OBC).
This method has high accuracy even near the disordered point.
In addition,
small size data are important
since we need to investigate poles far from the real axis.
We only need to relate the energy eigen values to the static structure factor.

In the next section,
we will discuss the relation between the static structure factor 
and the energy eigen values under OBC, 
according to the S-A prescription.\cite{Sorensen-Affleck}

\section{Edge states and Green function\label{sec_EG}}
In this section,
we discuss a Green function 
based on the S-A
prescription\cite{Sorensen-Affleck} [see Appendix \ref{app_a} in detail].

\subsection{Modified S-A prescription}
Now we consider a Green function $G(q,\kappa)$
which is the Fourier transform of $G(x,\tau)$ in Euclidean space-time.
The Green function determines
various physical quantities,
which contain
a static structure factor $S(q)$ and an energy gap of edge states. 
Between the Green function $G(q,\kappa)$
and a frequency $\omega_q$ (or an energy of a boson particle with
a wave number $q$),
the following relation is given in Appendix \ref{app_a}:
\begin{equation}
G(q,\kappa)=\frac{1}{\kappa^2 + \omega_q^2},
\end{equation}
where $\kappa$ is a imaginary frequency.
The static structure factor is obtained by applying the Fourier transform 
of $G(q,\kappa)$,
and then limiting as $\tau \to 0$;
\begin{subequations}
\label{eq_SQCORR}
\begin{align}
S(q)&=\lim_{\tau\to 0} \int \frac{d\kappa}{2\pi} 
G(q,\kappa)e^{i\kappa\tau}
=\frac{1}{2\omega_q},
\label{SQ}
\intertext{which recalls the original relation.
One can show the correlation function from the Fourier transform 
of the static structure factor.}
\langle {\bf S}_{x}\cdot {\bf S}_{y} \rangle
&\equiv
\int \frac{dq}{2\pi} e^{iq(x-y)}S(q).\label{CORR}
\end{align}
\end{subequations}

Next,
we shall examine the relation between the Green function and edge states.
The edge states mean the triplet states and the singlet state
under OBC.
Among these states,
there is
a energy difference:
\begin{equation}
 \Delta E_{\rm ST} \equiv E_{\rm triplet} - E_{\rm singlet}.
\end{equation}
The energy gap of edge states is connected with the Green function 
by the path integral method.
The details are given in Appendix \ref{app_a}.
Here, we only show
the relation between the energy gap of edge states and the Green function:
\begin{equation}
S_{\rm eff} =(-1)^{L}\lambda^2 {\bf S'}_1 \cdot {\bf S'}_L
\int d\tau_1 d\tau_L \frac{d\kappa dq}{(2\pi)^2} 
G(q,\kappa)e^{iq(L-1) +i\kappa(\tau_L - \tau_1)} \label{eq_Seff},
\end{equation}
where the left hand side of Eq. \eqref{eq_Seff} means
an effective action which is associated with an effective Hamiltonian,
$S_{\rm eff}=\int d\tau {\cal H}_{\rm eff}$,
and $\lambda$ is an interaction parameter between the $S=1/2$ 
edge spins ${\bf S'}_{1,L}$ and neighboring fields $\boldsymbol{\phi}$. 
The integral over $\tau_1$ or $\tau_L$ 
provides a factor of $\delta(\kappa)$.
Thus we obtain\cite{note-ST}
\begin{align}
\Delta E_{\rm ST} (L-1)&= (-1)^{L}\lambda^2 
\int \frac{dq}{2\pi} G(q,\kappa=0)e^{iq(L-1)}.
\label{DEST}
\end{align}
Comparing Eq. \eqref{eq_SQCORR} with Eq. \eqref{DEST},
we see that
Eq. \eqref{DEST} is more manageable. 
The reason is that
the integrand of Eq. \eqref{DEST},
{\it i.e.} the Green function,
has poles,
while
that of Eq. \eqref{eq_SQCORR}, 
{\it i.e.} the static structure factor, has branch points.

\subsection{From static structure factor to Green function}
In Sec. \ref{sec_SSF},
we have discussed the static structure factor.
We can apply a similar discussion to the Green function.
Corresponding to the static structure factor,
the Green function is permitted to have the following functional forms:
\begin{alignat}{2}
G_{\rm sing}(q,\kappa)&=
\frac{1}{\kappa^2 +
(2A)^{-2}((q+\widetilde{m}i)^2-d)((q-\widetilde{m}i)^2-d)},&\qquad&
 \text{({\it cf.} Eq. \eqref{eq_prod})} \label{eq_prod_g}
\end{alignat}
or
\begin{subequations}
\label{eq_diff_g}
\begin{alignat}{2}
G_{\rm sing}(q, \kappa)&=G^{+}_{\rm sing}(q,\kappa) 
+ G^{-}_{\rm sing}(q,\kappa), &\qquad&
 \text{({\it cf.} Eq. \eqref{eq_diff})}\\
\intertext{where}
G^{\pm}_{\rm sing}(q,\kappa)&=
\frac{1}{\kappa^2 - 
(A /\widetilde{m})^{-2}((q \mp \widetilde{m}i)^2-d)}.
& &
\end{alignat}
\end{subequations}
$G^{+}_{\rm sing}(q,\kappa)$ 
($G^{-}_{\rm sing}(q,\kappa)$) has the singularities only in  the upper
(lower) half $q$-plane.
They satisfy 
\begin{subequations}
\begin{align}
\overline{G^{\pm}_{\rm sing}(\overline{q},\overline{\kappa})} 
&= G^{\mp}_{\rm sing}(q,\kappa),\\
G^{\pm}_{\rm sing}(-q,\kappa) &= G^{\mp}_{\rm sing}(q,\kappa),\\
\intertext{and}
G^{\pm}_{\rm sing}(q, - \kappa) &= G^{\pm}_{\rm sing} (q,\kappa).
\end{align}
\end{subequations}
In Appendix \ref{app_b},
we show that the static structure factor (Eq. \eqref{eq_diff}) is deduced from
Eq. \eqref{eq_diff_g}.

As well as the static structure factor,
the Green function $G(q,\kappa)$ 
(both Eqs. \eqref{eq_prod_g} and \eqref{eq_diff_g})
must satisfy the following physical requirements (PRII):
\begin{itemize}
\item[1.] [reality on the real axes] 
$G(q,\kappa)=\overline{G(\overline{q},\overline{\kappa})}$.
\item[2.] [parity]
$G(q, \kappa) = G(-q, \kappa)$,
$G(q, \kappa) = G(q, -\kappa)$
\item[3.] [algebraic singularity]
$G(q,\kappa)$ is an analytic function of complex variables $q$ and $\kappa$
except for several poles.
\item[4.] [analyticity on the real axes]
Poles must not cross the real $q$ and $\kappa$
axes.
\item[5.] [$\alpha$-dependency near $\alpha_{\rm D}$]
$G(q,\kappa)$ is an analytic function of a real parameter $\alpha$
in the neighborhood of $\alpha_{\rm D}$.
\end{itemize}
However,
Eq. \eqref{eq_prod_g} is different from Eq. \eqref{eq_diff_g} when $d=0$
while 
the static structure factors (Eqs. \eqref{eq_prod} and \eqref{eq_diff}) 
are the same ({\it cf.} PRI-6 in Sec. \ref{subsec_ssf}).
Another difference is that in the limit $q \to \infty$
Eq. \eqref{eq_prod_g} behaves as $q^{-4}$
while Eq. \eqref{eq_diff_g} as $q^{-2}$.
Hence,
it is easier to distinguish Eqs. \eqref{eq_prod_g}
and \eqref{eq_diff_g} 
clearer
than
Eqs. \eqref{eq_prod} and \eqref{eq_diff} near the disordered point
$\alpha=\alpha_{\rm D}$.

\subsection{energy gap of edge states}

On the basis of the above discussion with Eq. \eqref{DEST},
the energy gap of edge states obtained from Eq. \eqref{eq_prod_g} is
\begin{subequations}
\label{eq_Eprod}
\begin{align}
\Delta E_{\rm ST} (L-1) &=(-1)^{L}\lambda^2 
\frac{A^2 e^{-\widetilde{m}(L-1)}}{\widetilde{m}\sqrt{d}\sqrt{\widetilde{m}^2+d}}
\sin(\sqrt{d}(L-1) + \phi(\widetilde{m},d)), \nonumber\\
&=(-1)^{L}\widetilde{A}e^{-\widetilde{m}(L-1)}\sin(\sqrt{d}(L-1)+\phi(\widetilde{m},d)),
\label{eq_Eproda}
\intertext{for $d>0$ (or $\alpha>\alpha_{\rm D}$),
where $\phi(\widetilde{m},d)=\tan^{-1}(\sqrt{d}/\widetilde{m})$,
and also}
\Delta E_{\rm ST} (L-1) &=(-1)^{L}\lambda^2 
\frac{A^2 e^{-\widetilde{m}(L-1)}}
{\widetilde{m}\sqrt{-d}\sqrt{\widetilde{m}^2+d}}
\sinh(\sqrt{-d}(L-1)+\phi(\widetilde{m},d)), \nonumber\\
&=
(-1)^{L}\widetilde{A}e^{-\widetilde{m}(L-1)}\sinh(\sqrt{-d}(L-1)+\phi(\widetilde{m},d)),
\label{eq_Eprodb}
\end{align}
\end{subequations}
for $d<0$ (or $\alpha<\alpha_{\rm D}$), where 
$\phi(\widetilde{m},d)=\tanh^{-1}(\sqrt{-d}/\widetilde{m})$.

On the other hand, 
the energy gap of edge states about Eq. \eqref{eq_diff_g} is
\begin{subequations}
\label{eq_Ediff}
\begin{align}
\Delta E_{\rm ST}(L-1)
&=(-1)^{L}\lambda^2 
\frac{A^2 e^{-\widetilde{m}(L-1)}}
{\widetilde{m}^2\sqrt{d}}
\sin(\sqrt{d}(L-1)), \nonumber\\
&=
(-1)^{L}\widetilde{A}e^{-\widetilde{m}(L-1)}\sin(\sqrt{d}(L-1)),
\label{eq_Ediffa}
\intertext{
for $\alpha>\alpha_{\rm D}$, and also}
\Delta E_{\rm ST}(L-1)
&=(-1)^{L}\lambda^2 
\frac{A^2 e^{-\widetilde{m}(L-1)}}
{\widetilde{m}^2\sqrt{-d}}
\sinh(\sqrt{-d}(L-1)),\nonumber\\
&=
(-1)^{L}\widetilde{A}e^{-\widetilde{m}(L-1)}\sinh(\sqrt{-d}(L-1)),
\label{eq_Ediffb}
\end{align}
\end{subequations}
for $\alpha<\alpha_{\rm D}$.

We will verify which is more appropriate between these two predictions
(Eqs. \eqref{eq_Eprod} and \eqref{eq_Ediff})
by analyzing numerical data
in Sec. \ref{sec_NA}.
Note that 
Eq. \eqref{eq_Eprod} is apparently different from Eq. \eqref{eq_Ediff} when $L=1$:
Eq. \eqref{eq_Ediff} is always equal to zero,
whereas Eq. \eqref{eq_Eprod} is nonzero.

\section{Implementation for lattice \label{sec_LI}}
In this section,
we consider an effect of the lattice structure. 
Equations \eqref{eq_Eprod} and \eqref{eq_Ediff} are not equal to zero
even when $L$ is not an integer number,
and therefore
they are incompatible with the lattice structure.
To include the lattice structure,
we must require $S(q)=S(q+2\pi)$ and $G(q,\kappa)=G(q+2\pi,\kappa)$.

Now, we organize the new physical requirements (PRIII)
for the static structure factor and the Green function,
considering the lattice structure.
PRIII from 1 to 5 are the same as 
PRI and PRII.
We add the requirement of the periodicity to PRIII:
\begin{itemize}
\item[6.] [periodicity] $S(q)=S(q+2\pi)$ and $G(q,\kappa)=G(q,\kappa+2\pi)$.
\end{itemize}
From PRIII-6, 
we derive another physical requirement:
\begin{itemize}
\item[7.] [singularity in the Brillouin zone] 
	  There are only four singular points (poles or algebraic
	  singularity)
	  in the first Brillouin zone ($-\pi < \Re q \le \pi$).
\end{itemize}
All the information needed for any problem can be determined 
in this zone.

Then, we can construct some static structure factors and Green functions,
satisfying these requirements,
and we show them in Secs. \ref{subsec_ISV} and \ref{subsec_SWV}.

\subsection{Infinite sum version\label{subsec_ISV}}
The easiest way is to consider 
the infinite sum of the translated singular parts.
The static structure factor has the form as
\begin{subequations}
\label{eq_infinitesum}
\begin{align}
S(q)&=\sum_{j=-\infty}^{\infty} S_{\rm sing}(q+2\pi j) +S_{\rm reg}(q),
\label{eq_infinitesumS}\\
\intertext{and the Green function has}
G(q,\kappa)&=\sum_{j=-\infty}^{\infty}G_{\rm sing}(q+2\pi j,\kappa) +
 G_{\rm reg}(q,\kappa), \label{eq_infinitesumG}
\end{align}
\end{subequations}
where both $S_{\rm sing}(q + 2\pi j)$ and $G_{\rm sing}(q + 2\pi j)$
represent shifted singular terms.
$S_{\rm reg}(q)$ and $G_{\rm reg}(q, \kappa)$ are 
regular functions in the whole $q$ plane,
such as
\begin{equation}
S_{\rm reg} (q) = \sum_{l=0}^{\infty} a_l \cos(l q),
\end{equation}
where $a_l$ is a real number.
In Eq. \eqref{eq_infinitesum},
the singular terms
correspond to long-range behaviors in the real space,
whereas
the regular terms correspond
to model-dependent short-range behaviors.

Note that
the infinite sum \eqref{eq_infinitesum}
for Eq. \eqref{eq_ssf_com} or Eq. \eqref{eq_ssf_incom}
is divergent,
whereas that for Eq. \eqref{eq_prod} or Eq. \eqref{eq_diff} is convergent.

\subsection{Sine wave version\label{subsec_SWV}}
Alternatively,
substituting a $2\pi$ or $4\pi$-periodic function $p(q)$
for $q$ in $S(q)$ or $G(q, \kappa)$,
we also obtain a $2\pi$-periodic static structure factor $S(p(q))$ 
or a $2\pi$-periodic Green function $G(p(q), \kappa)$, respectively.
We impose some constraints on the periodic function $p(q)$ to satisfy PRIII:
\begin{itemize}
\item[1.] $p(q)$ is a holomorphic function.
\item[2.] $p(q+2\pi)=p(q)$ or $p(q+2\pi)=-p(q)$.
\item[3.] $p(-q)=p(q)$ or $p(-q)=-p(q)$.
\item[4.] $\overline{p(\overline{q})}=p(q)$ or $\overline{p(\overline{q})}=-p(q)$.
\item[5.] The inverse function $p(q)^{-1}$ is a single-valued function
in the first Brillouin zone $-\pi < \Re q \le \pi$.
\item[6.] $\lim_{q\to 0} p(q)/q=1$.
\end{itemize}

The above requirements
determine the distribution of zeros of $p(q)$.
From Weierstrass's theorem for infinite products\cite{Ahlfors}
and the above constraints,
the function $p(q)$
is determined as
\begin{equation}
p(q)=2\sin \frac{q}{2}.
\end{equation}

Replacing $q$ in $S(q)$ and $G(q,\kappa)$ by $p(q)$,
the static structure factor can be described as
\begin{subequations}
\label{eq_sine}
\begin{align}
S(q)&=S_{\rm sing}(p(q)) +S_{\rm reg}(q),\label{eq_sineS}\\
\intertext{and the Green function as}
G(q,\kappa)&=G_{\rm sing}(p(q),\kappa) +
 G_{\rm reg}(q,\kappa).\label{eq_sineG}
\end{align}
\end{subequations}

\subsection{Contour}

\begin{figure}[h]
\includegraphics[width=8cm]{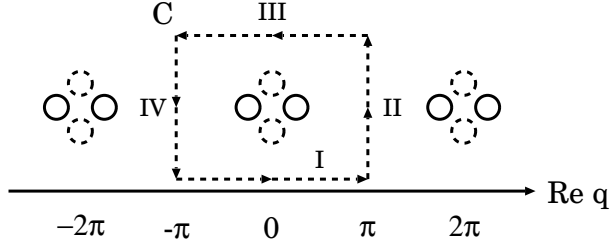}
\caption{Contour $\rm C$ for the integral over $q$ 
in Eqs. \eqref{eq_SQCORR} and \eqref{DEST}. }
\label{fig_contour}
\end{figure}

Corresponding to both the infinite sum version and the sine wave version,
the contour $\rm C$
of the integral over $q$ in Eqs. \eqref{eq_SQCORR} and \eqref{DEST}
is described in
Fig. \ref{fig_contour}.
Solid circles mean 
 poles or branches for the incommensurate case, 
and broken circles for the
 commensurate case.
$\rm I$, $\rm II$, $\rm III$, and $\rm IV$ 
represent the contours
$\{q| (\Re q: -\pi \to \pi) \cap (\Im q = 0)\}$, 
$\{q| (\Re q =\pi) \cap (\Im q: 0\to \infty)\}$,
$\{q| (\Re q:\pi \to -\pi) \cap (\Im q = \infty)\}$,
and $\{q| (\Re q=-\pi )\cap (\Im q : \infty \to 0) \}$,
respectively.
The contributions of $\rm II$ and ${\rm IV}$  cancel each other out
because of the periodicity.
The contribution of $\rm III$ can be ignored since
$S(q)$ and $G(q,\kappa) \lesssim q^{-2}$ as $q \to \infty$.
We, therefore, obtain
that
$\oint_{\rm C}=\int_{\rm I}$.
As a result,
the integral of the infinite sum Green function
(Eq. \eqref{eq_infinitesumG})
is equal to Eqs. \eqref{eq_Eprod} and \eqref{eq_Ediff}.
A similar discussion can be applied to 
the static structure factor.

Note that the integral of the sine wave Green function
 (Eq. \eqref{eq_sineG})
is different from
Eqs. \eqref{eq_Eprod} and \eqref{eq_Ediff}.
We consider it in detail in Appendix \ref{app_c}.

\section{Numerical analysis\label{sec_NA}}
Our aim in this study is to decide 
between Eqs. \eqref{eq_prod} and \eqref{eq_diff}.
In the previous section,
each behavior of the energy gap of edge states has been expected from
Eq. \eqref{eq_prod} or \eqref{eq_diff}.
In this section,
we, therefore, carry out the numerical calculation of the energy gap
between the triplet and singlet states,
and verify
whether the results correspond to the predictions
(Eqs. \eqref{eq_Eprod} and \eqref{eq_Ediff})
with the use of a nonlinear least-squares (NLLS) fitting program,
which needs appropriate initial values.
Applying the previous results,\cite{Nomura-Murashima}
we guess the initial values first.

Although we have calculated the incommensurate wave number $q_{\rm IC}$ 
in Ref. [\onlinecite{Nomura-Murashima}],
its analytical reasoning
 was unclear.
Also,
we have not so far investigated $\widetilde{m}$ 
(the distance between the real axis
and the center of two singular points)
and $\widetilde{A}$
(the amplitude of the energy gap).
On the basis of the S-A prescription,
we will calculate them in this section.
We will also trace the singularities in the commensurate region.

\subsection{Surveys of edge states and incommensurability}
We treat the $S=1$ BLBQ chain under OBC,
\begin{subequations}
\label{eq_BLBQHamiltonian}
\begin{align}
 {\cal H} &= \sum_{i=1}^N h_i ,\\
 h_i &= {\bf S}_i \cdot {\bf S}_{i+1} 
+ \alpha ({\bf S}_i\cdot{\bf S}_{i+1}),
\end{align}
\end{subequations}
where $N$ is the number of the sub-Hamiltonian $h_i$
and $\alpha$ is the interaction constant of the biquadratic term.

Note that $N=L-1$, where $L=\{6, 7, \cdots, 14\}$ is the chain length.
We can treat long chains ($L > 14$).
However, their significant digit is smaller 
than that of short chains ($L\le 14$)
since their amplitudes of the energy gap are exponentially small near the AKLT
point.
Thus we treat up to $L=14$.
We exclude
data smaller than $L=6$
since the short-range behaviors are affected
by model-dependent regular terms,
{\it i.e.} $S_{\rm reg}(q)$ and $G_{\rm reg}(q, \kappa)$
in Eqs. \eqref{eq_infinitesum} and \eqref{eq_sine}

Since there are two edge $S=1/2$ spin freedoms at the AKLT point
($\alpha = \alpha_{\rm D}$),
the following degeneracy occurs:
\begin{equation}
(S=1/2)\otimes(S=1/2)=(S=0)\oplus(S=1),
\end{equation}
which reflects the $Z_2 \times Z_2$ symmetry.
\cite{Kennedy,Hagiwara-Katsumata,Glarum-Geschwind,Mitra}
Therefore, the singlet-triplet energy gap (or the gap of edge states)
\begin{equation}
\Delta E_{\rm ST}(N,\alpha) \equiv
E_{\rm triplet}(N,\alpha) - E_{\rm singlet}(N,\alpha),
\end{equation}
is zero for all length spin chains at the AKLT point:
\begin{equation}
\Delta E_{\rm ST}(N,\alpha_{\rm D}) = 0.
\end{equation}
Note that in the thermodynamic limit
the triplet states and the singlet state also become degenerate in the
whole Haldane phase ($-1<\alpha<1$), 
and thus
the amplitude of the gap of edge states goes to zero as $N\to \infty$:
\begin{equation}
\lim_{N\to \infty} \Delta E_{\rm ST}(N,\alpha)=0.
\end{equation}
To avoid confusion,
we call the degeneracy at the AKLT point as the AKLT degeneracy.

\begin{figure}[h]
 \includegraphics[width=8.0cm]{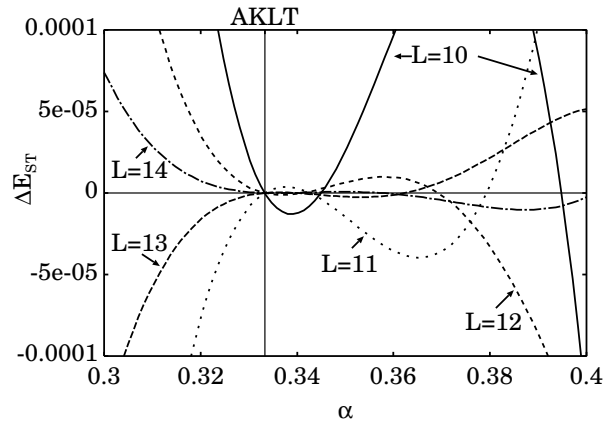}%
 \caption{Energy gaps of edge states
$\Delta E_{\rm ST} = E_{\rm triplet} -E_{\rm singlet}$ 
as a function of $\alpha$ for various sizes.}
\label{fig_EST}
 \end{figure}

Numerical results of the gap of edge states are shown in Fig. \ref{fig_EST}.
For $\alpha \ne \alpha_{\rm D}$, the AKLT degeneracy breaks down.
We see oscillating behaviors in the gap of edge states for $\alpha >
\alpha_{\rm D}$.
This phenomenon has been predicted from Eqs. \eqref{eq_Eproda} and \eqref{eq_Ediffa}.
Note that
for $\alpha <\alpha_{\rm D}$ the sign of the gap of edge states is different
between
even length chains and odd length chains because the parity of the bulk
is different among these chains.\cite{Sorensen-Affleck}

\subsection{Initial guess}

\subsubsection{incommensurate wave number}
\begin{figure}[h]
 \includegraphics[width=8.0cm]{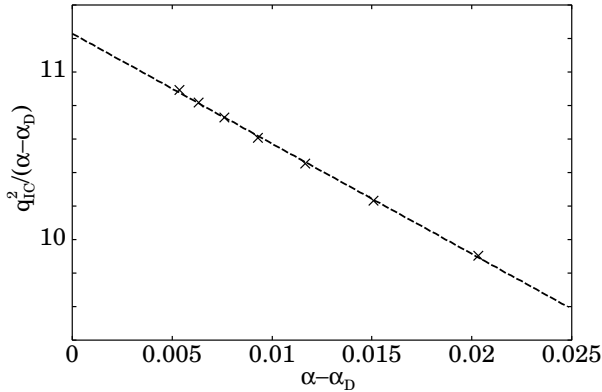}%
 \caption{Dependence of the incommensurate wave number $q_{\rm IC}$
on $\alpha-\alpha_{\rm D}$.
\label{fig_q2}}
\end{figure}

Since the gap of edge states $\Delta E_{\rm ST}$ is a function of
$\alpha$ and $N$, and it is oscillating in the incommensurate phase, 
we can find the relation between $\alpha$ and $N$,
taking account of the condition $\Delta E_{\rm ST} = 0$.
Then we consider the $n$-th zero point of the singlet-triplet gap,
\begin{equation}
\Delta E_{\rm ST}(N,\alpha_n(N))=0.
\end{equation}
If we adopt Eq. \eqref{eq_Ediffa}, 
{\it i.e.} $\Delta E_{\rm ST}(N) \sim \sin(q_{\rm IC} N)$, 
in the incommensurate region,
we can relate the incommensurate wave number $q_{\rm IC}$ with $N$ as
\begin{equation}
q_{\rm IC}(\alpha_n (N)) = \frac{\pi n}{N}, \label{eq_qIC}
\end{equation}
where $n=1, 2, 3, \cdots$.
We have already found that
they are fitted by a universal curve like 
$\sqrt{\alpha - \alpha_{\rm D}}$.\cite{Nomura-Murashima}
We show $q_{\rm IC}^2/(\alpha - \alpha_{\rm D})$ as a function of
$\alpha - \alpha_{\rm D}$ in Fig. \ref{fig_q2}.
These data fit well with the following equation:
\begin{equation}
d(\alpha) \equiv q_{\rm IC}^2 = d_1 (\alpha - \alpha_{\rm D}) 
+ d_2 (\alpha - \alpha_{\rm D})^2 + O((\alpha-\alpha_{\rm D})^3), \label{eq_qic}
\end{equation}
where $d_1 = 11.230 \pm 0.010$ and $d_2 = -65.76 \pm 0.83$.

If we adopt Eq. \eqref{eq_Eproda},
the corrections of $O(1/N)$ in $d_1$ and $d_2$ should be found.
Since
their corrections are smaller than 2 $\%$,
we see that our guess adopting Eq. \eqref{eq_Ediffa}, 
$\Delta E_{\rm ST}(N) \sim
\sin(\sqrt{d}N)$, is more reliable than Eq. \eqref{eq_Eproda}.
We expect that the gap of edge states behaves in the incommensurate and
commensurate
regions
as Eq. \eqref{eq_Ediffa} and Eq. \eqref{eq_Ediffb},
respectively.
This result means that
the static structure factor is Eq. \eqref{eq_diff}.

In addition,
the number of zero points $n$ (except the AKLT point)
and the system size $N$($=\{1,\cdots,13\}$)
are correlative.
It is easy to find the following relation:
\begin{equation}
n \equiv N \pmod{3}.
\end{equation}
We see that the relation between the max number of zero
points $n^{\rm max}$ and $N$ as
\begin{equation}
\frac{\pi n^{\rm max}(N)}{N}< \frac{\pi}{3}.
\end{equation}
We can confirm this relation up to $N=13$ ($n^{\rm max}(13)=4$).
We expect that $\pi n^{\rm max}(N)/N$ has the limit $\pi/3$ 
as $N$ tends to $\infty$,
and therefore
the position of the max-$n$-th zero point $\alpha_{n^{\rm max}}$ goes to the ULS point
as $N\to\infty$.

\subsubsection{amplitude and center of coupling poles}
In the previous sub-subsection 
we have found 
that the Green function
corresponds to the difference type of 
the static structure factor (Eq. \eqref{eq_diff}).
Next, we examine the parameter $\widetilde{A}$ and $\widetilde{m}$.

\begin{figure}[h]
\includegraphics[width=8cm]{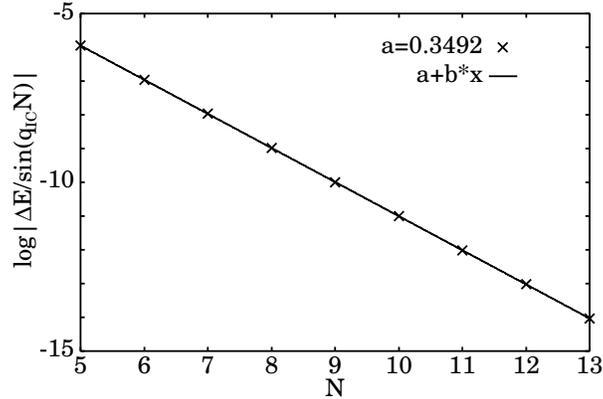}
\caption{Finite size results for
$\log |\Delta E_{\rm ST}/\sin(q_{\rm IC}N)|$ 
when $\alpha = 0.3492$.}
\label{fig_log_ic}
\end{figure}

In the incommensurate region $\alpha >\alpha_{\rm D}$,
we expect 
that the gap of edge states has the
following form:
\begin{equation}
\Delta E_{\rm ST}(N)=(-1)^{N+1}\widetilde{A}e^{-\widetilde{m}N}\sin(q_{\rm IC}N).
\end{equation}
The incommensurate wave number $q_{\rm IC}(\alpha)$ near the AKLT point
is obtained in the previous sub-subsection.
Using these values and considering the following equation,
we can determine $\widetilde{A}$ and $\widetilde{m}$:
\begin{equation}
 \log \left|\frac{\Delta E_{\rm ST}(N)}
{\sin(q_{\rm IC}N)}
\right|=\log |\widetilde{A}| -\widetilde{m}{N}. \label{eq_log_ic}
\end{equation}
Figure \ref{fig_log_ic} shows $\log|\Delta E_{\rm ST}/\sin(q_{\rm
IC}N)|$ when $\alpha=0.3492$
behaves linearly as a function of $N$.
We have just confirmed our prediction for the incommensurate
region.

Similar consideration can be applied to the commensurate region.
In the region,
the gap of edge states should be
\begin{equation}
\Delta E_{\rm ST}(N)=(-1)^{N+1}\widetilde{A}e^{-\widetilde{m}N}\sinh(q_{\rm C}N),
\end{equation}
where $q_{\rm C}=\sqrt{-d}=\sqrt{-(\alpha-\alpha_{\rm
D})}\sqrt{d_1 - d_2(\alpha-\alpha_{\rm D})}$.
Here,
the commensurate wave number $q_{\rm C}$ 
is indirectly
determined
 by using $d_1$ and $d_2$,
which are obtained from Eq. \eqref{eq_qic}.
Considering the following equation,
we can obtain $\widetilde{A}$ and $\widetilde{m}$ in the commensurate region:
\begin{equation}
\log \left|\frac{\Delta E_{\rm ST}(N)}{\sinh(q_{\rm C}N)}\right|
=\log|\widetilde{A}|-\widetilde{m}N.
\end{equation}

\subsection{Nonlinear least-squares fitting }
In the previous subsection,
we have adopted the commensurate wave number $q_{\rm C}$ 
indirectly determined by using
the parameters of
the incommensurate wave number in Eq. \eqref{eq_qic},
although the relation between $q_{\rm IC}$ and $q_{\rm C}$ is somewhat
unclear.
Actually,
it seems that
the region, 
where 
Eq. \eqref{eq_qic} is permitted,
may be
narrower in the commensurate side
than in the incommensurate side.
In order to check the above mentioned results from another viewpoint,
we employ a NLLS fitting method.
Using this method,
we can determine above parameters 
$\widetilde{A}$, $\widetilde{m}$,
and $d$ directly.
Since the method requires appropriate initial values,
we must have determined them in the previous subsection.

Taking into account the fact that
the amplitude of the energy gap is exponentially small 
near the AKLT point,
we use the following weighted values to perform the NLLS
fitting program:
\begin{equation}
y_N = (-1)^{N+1}\Delta E_{\rm ST}(N)w_N,
\end{equation}
where $w_N\equiv\exp(\widetilde{m}'N)$ is a weight, 
and $\widetilde{m}'=\widetilde{m}+\delta$ is a 
value estimated from $\widetilde{m}$ at the nearest $\alpha$.
Correctly,
what we determine by the NLLS fitting method
is not $\widetilde{m}$ but $\delta$.

The NLLS fitting method requires the minimization of the squared residuals,
\begin{equation}
Q = \sum_{N}\frac{1}{w_N^2}(y_N-f_N(\hat{\bf x}))^2, 
\end{equation}
where $\hat{\bf x}\equiv (\widetilde{A}, \widetilde{m}, d)$
and $f_N(\hat{\bf x})$ is a fitting function of $\hat{\bf x}$.
From the minimum value of $Q$,
we obtain parameters $(\widetilde{A}, \widetilde{m}, d)$.

\begin{figure}[ht]
\includegraphics[width=8.0cm]{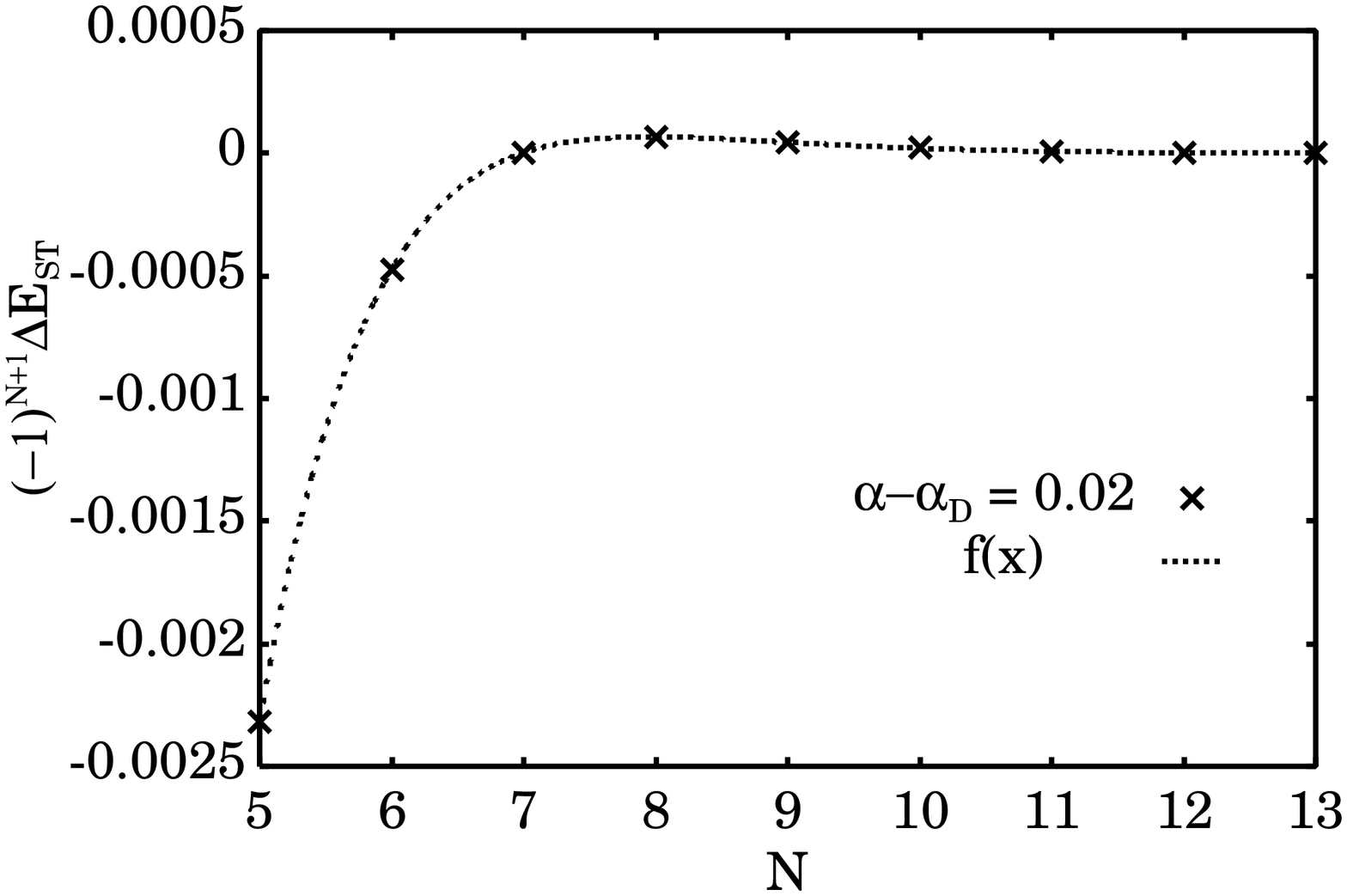}
\caption{Fitting results with Eq. \eqref{eq_Ediff} 
when $\alpha-\alpha_{\rm D}=0.02$.}
\label{fig_NLS0d02}
\end{figure}

\begin{figure}[ht]
\includegraphics[width=8.0cm]{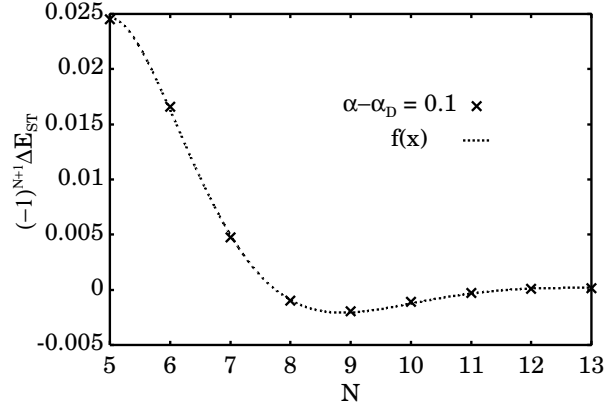}
\caption{Fitting results with Eq. \eqref{eq_Ediff} 
when $\alpha-\alpha_{\rm D}=0.1$.}
\label{fig_NLS0d1}
\end{figure}

In the case of $\alpha-\alpha_{\rm D}=0.02$,
for example,
we show the data of the energy gap
and the fitting function $f(\hat{\bf x})=\widetilde{A}\exp(-\widetilde{m}N)
\sin(\sqrt{d}N)$
where $\widetilde{A}=-0.421$, $\widetilde{m}=0.991$, and $d=0.202$
in Fig. \ref{fig_NLS0d02}.
Also,
the case of $\alpha - \alpha_{\rm D}=0.1$ is shown
in Fig. \ref{fig_NLS0d1}.
The parameters of fitting function $f(\hat{\bf x})$
are
$\widetilde{A}=-0.758$, $\widetilde{m}=0.638$, and $d=0.810$.

\subsubsection{fitting with Eq. \eqref{eq_Ediff}}

\begin{figure}[ht]
\includegraphics[width=8.0cm]{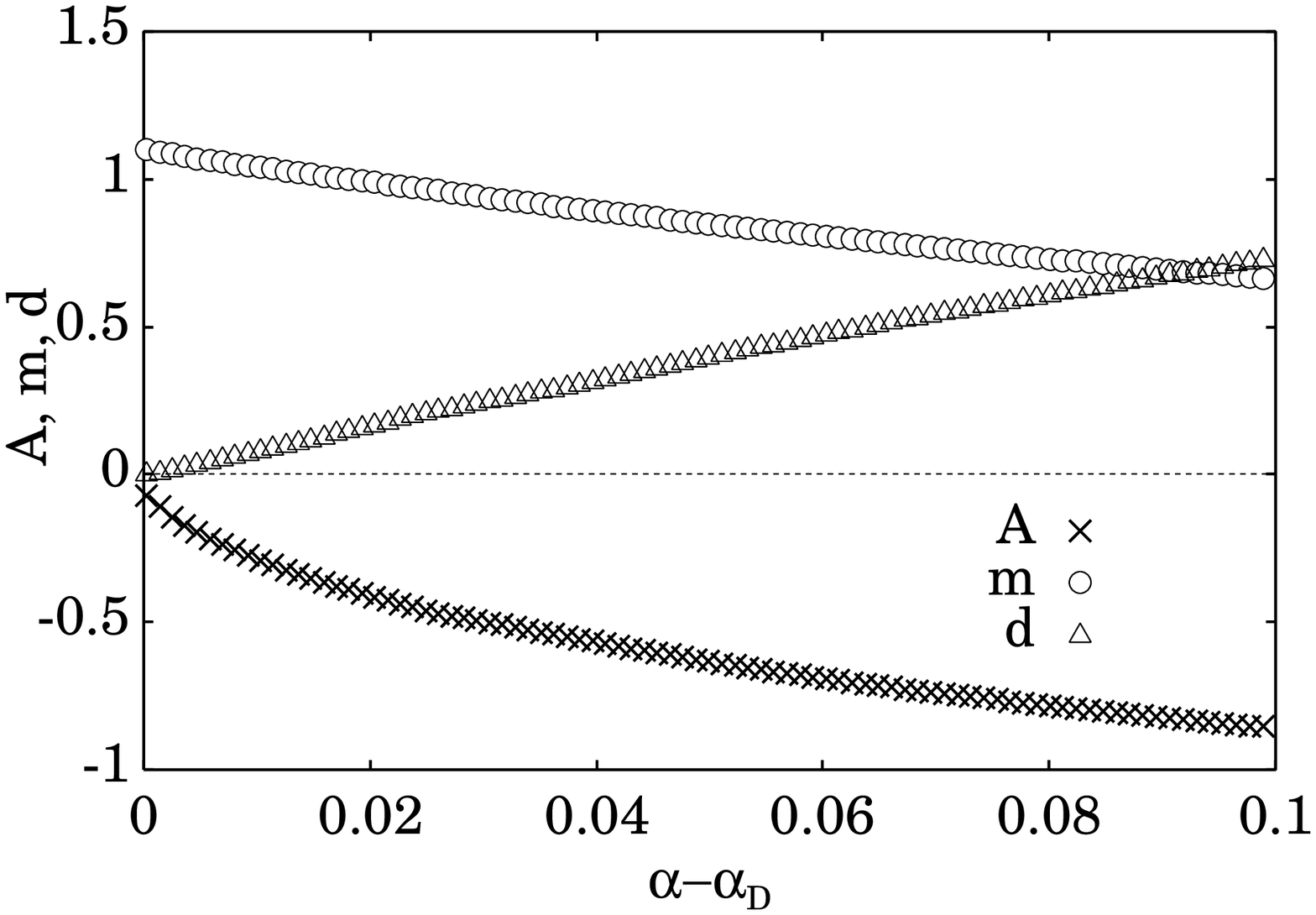}
\caption{Nonlinear least-squares fitting results with
 Eq. \eqref{eq_Ediff}.}
\label{fig_TT}
\end{figure}

\begin{figure}[ht]
 \includegraphics[width=8.0cm]{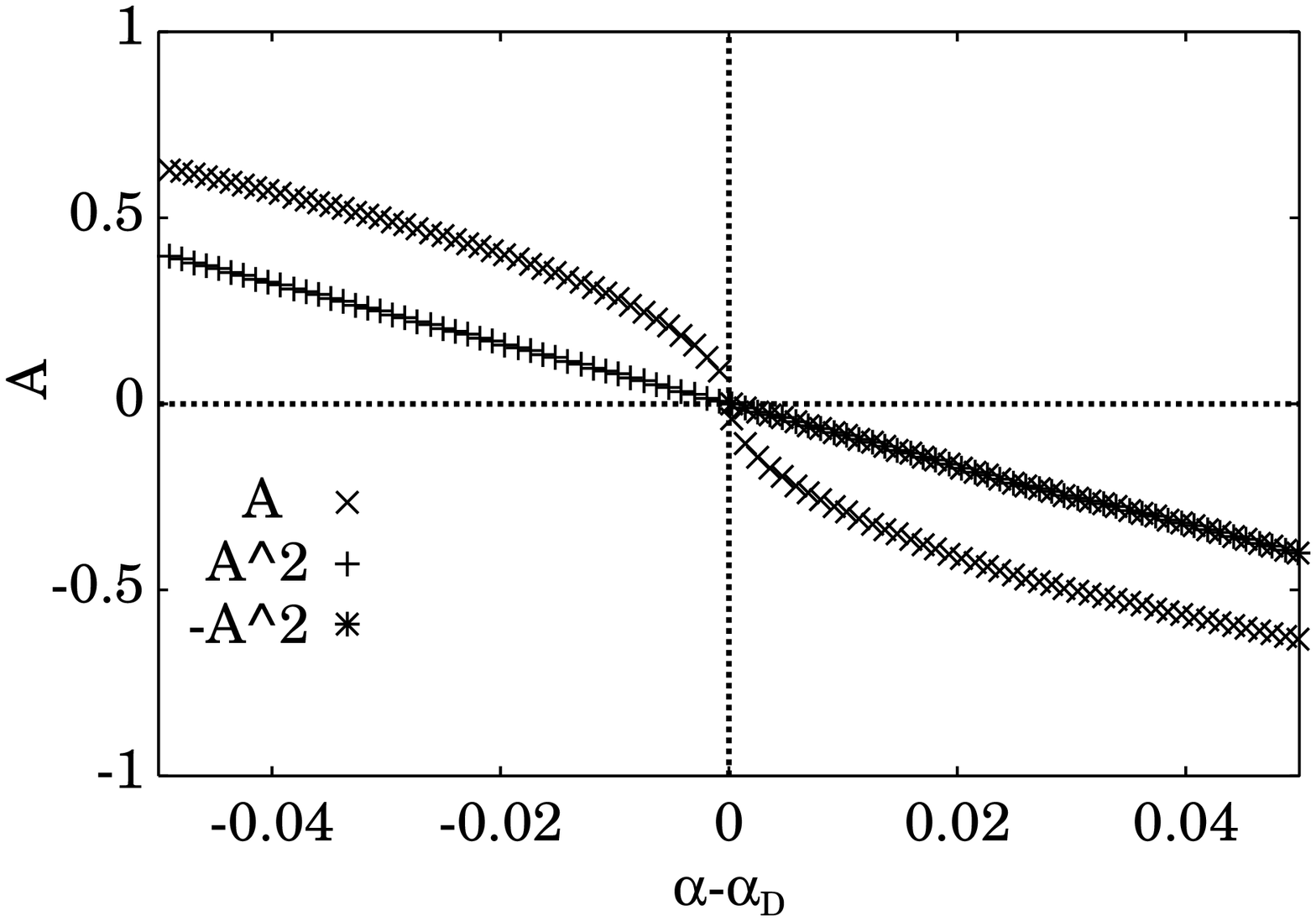}%
 \caption{Nonlinear least-squares fitting results for
  $\widetilde{A}$. 
\label{fig_nl_A}}
 \end{figure}

\begin{figure}[ht]
 \includegraphics[width=8.0cm]{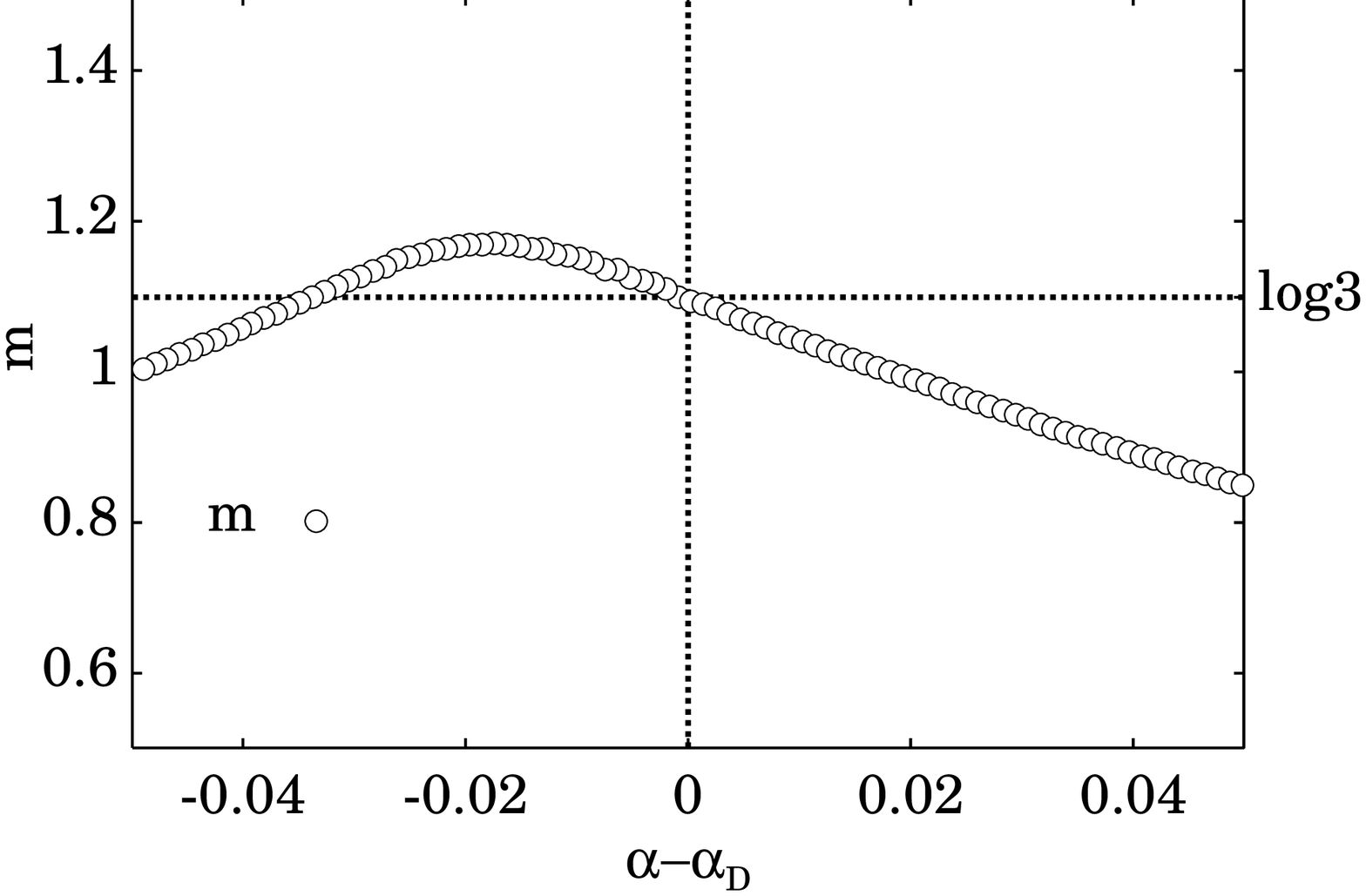}%
 \caption{Nonlinear least-squares fitting results for
  $\widetilde{m}$. 
\label{fig_nl_m}}
 \end{figure}

\begin{figure}[ht]
 \includegraphics[width=8.0cm]{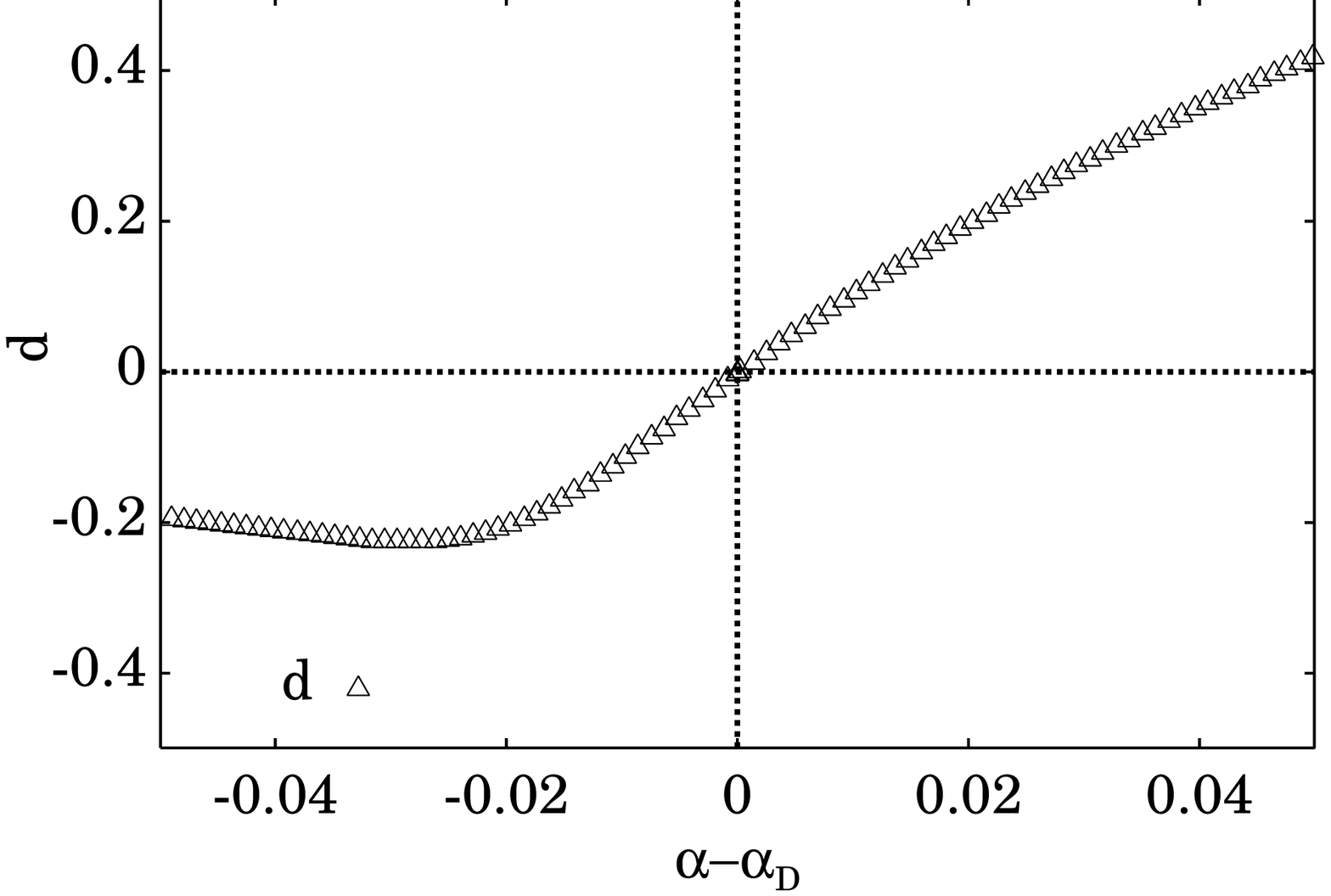}%
 \caption{Nonlinear least-squares fitting results for $d$. 
\label{fig_nl_q}}
 \end{figure}

Figure \ref{fig_TT} summarizes the fitting results with
Eq. \eqref{eq_Ediff}
in the incommensurate region.
The obtained parameters, $\widetilde{A}$, $\widetilde{m}$, and $d$
for $0\le$ $|\alpha - \alpha_{\rm D}|$ $\le 0.05$,
in which region $Q$ is less than  1.0 $\times$ $10^{-8}$,
are shown in Figs. \ref{fig_nl_A},  \ref{fig_nl_m}, and \ref{fig_nl_q},
respectively.
Near the AKLT point,
they converge very well,
and behave continuously with $\alpha$.

We see that these parameters are smooth between the commensurate
and incommensurate regions.
In Fig. \ref{fig_nl_A},
$\widetilde{A}^2$ is highly linear.
In Figs. \ref{fig_nl_m} and \ref{fig_nl_q},
each parameter $\widetilde{m}$, $d$
varies linearly in the incommensurate region,
whereas there are a broad maximum and a broad minimum,
respectively, 
at $\alpha-\alpha_{\rm D}=-0.02$ 
in the commensurate region.
The range
where
$d, \widetilde{m}$, and $A$ can be expanded 
in terms of $\alpha-\alpha_{\rm D}$
is narrower in the commensurate region than
in the incommensurate region.
When $\alpha-\alpha_{\rm D}$ is less than -0.02,
there should be a different mechanism 
from what we have expected in Sec. \ref{subsec_taylor},
since PRIII-5 is not satisfied in the region.

\begin{figure}[h]
\includegraphics[width=7.0cm]{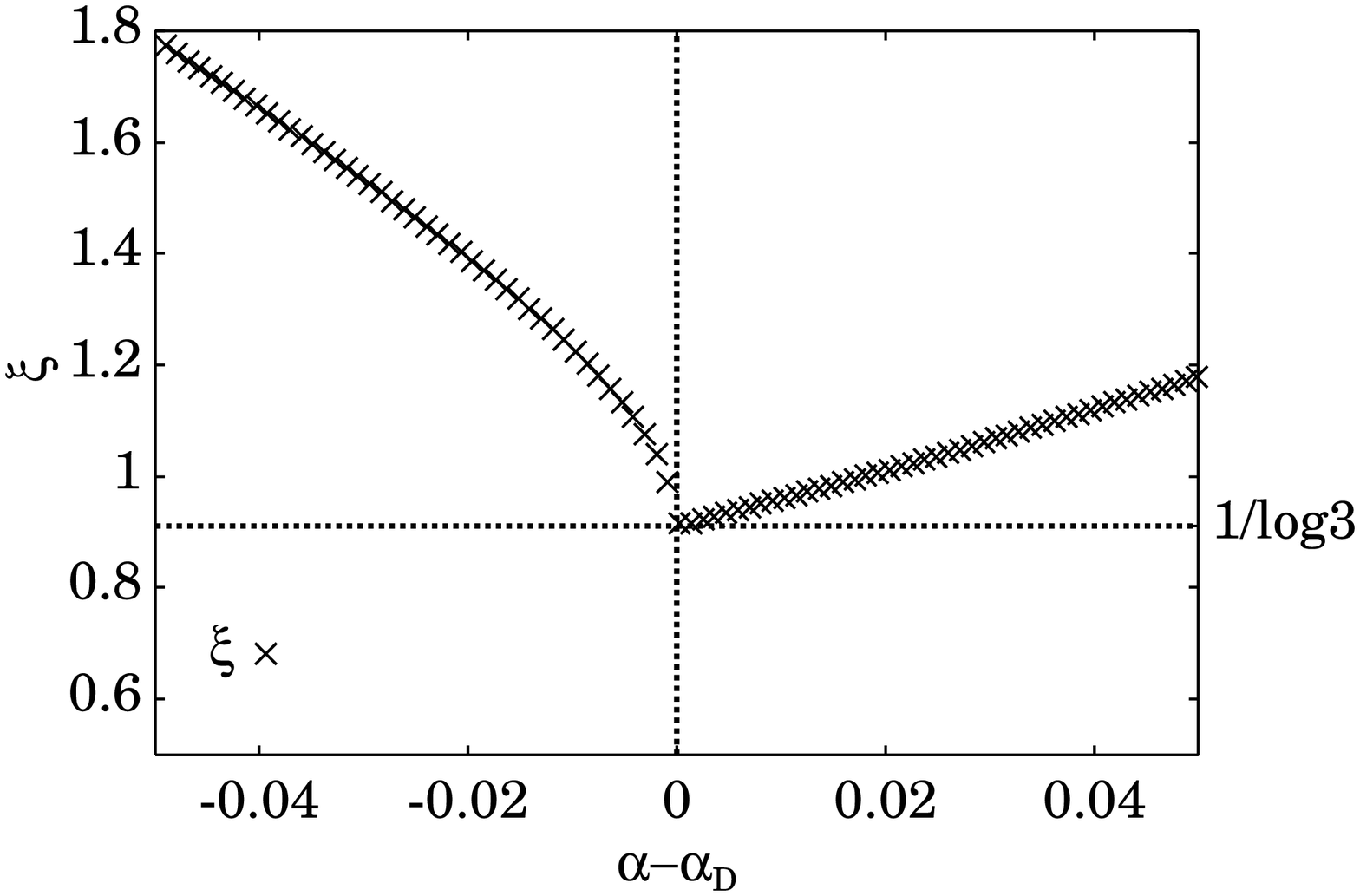}
\caption{Correlation length for $0\le|\alpha-\alpha_{\rm D}|\le 0.05$.}
\label{fig_correlation}
\end{figure}

Now, we estimate the correlation length $\xi$ from the obtained data.
Usually,
the correlation length is related to an inverse of a distance
between the closest singular point and the
real axis.
In the incommensurate region,
$\xi=\widetilde{m}^{-1}$, 
while $\xi=(\widetilde{m}-\sqrt{-d})^{-1}$ in the
commensurate region.
These results are shown in Fig. \ref{fig_correlation}.
This behavior is consistent with the previous
numerical
result.\cite{Schollwock-Jolicoeur-Garel}

\subsubsection{fitting with Eq. \eqref{eq_Eprod}}
\begin{figure}[ht]
 \includegraphics[width=8.0cm]{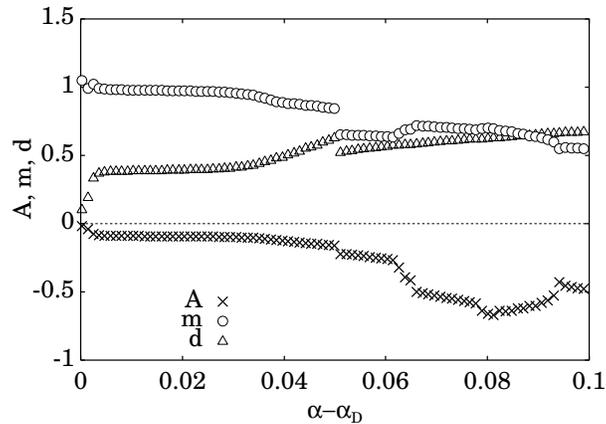}%
 \caption{Nonlinear least-squares fitting results with Eq. \eqref{eq_Eprod}.
\label{fig_NL_prod}}
 \end{figure}

We also attempt to apply the nonlinear least-squares fitting program 
to Eq. \eqref{eq_Eprod}.
The obtained parameters 
($\widetilde{A}$, $\widetilde{m}$, $d$) 
are shown in Fig. \ref{fig_NL_prod}.
We see that they behave as
discontinuous pieces about $\alpha$.
In addition,
the region
where $d \propto \alpha - \alpha_{\rm D}$ is very narrow.
These facts mean that
the supposed functions (Eq. \eqref{eq_prod} and
Eq. \eqref{eq_Eprod})
are not correct.
Of course, the residual $Q$ is larger than 
the one shown in the previous sub-subsection.

\subsubsection{about sine wave version}
\begin{figure}[ht]
\includegraphics[width=8.0cm]{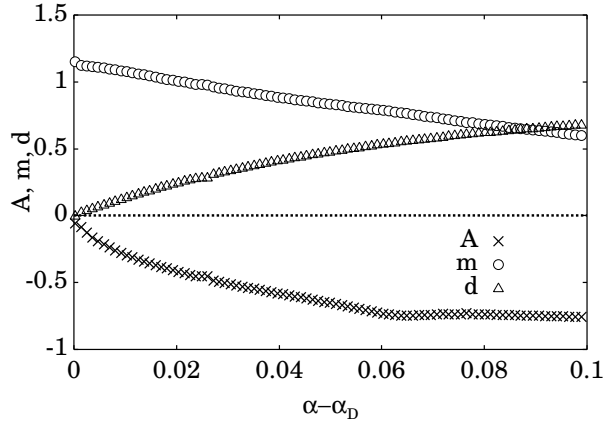}
\caption{Nonlinear least-squares fitting results with
 Eq. \eqref{eq_Ediffsine}.}
\label{fig_Ediffsine}
\end{figure}

We have so far assumed the infinite sum version (Eq. \eqref{eq_infinitesum})
as a lattice effect.
Now we consider the case of the sine wave version (Eq. \eqref{eq_sine}).
The energy gap of edge states (in
the incommensurate region)
is modified as 
\begin{equation}
\Delta E_{\rm ST}(L-1)=(-1)^{L}\widetilde{A} e^{\Re(\zeta) (L-1)}
\sin(\Im(\zeta)(L-1)), \qquad \text{({\it cf.} Eq. \eqref{eq_Ediffa})}
\label{eq_Ediffsine}
\end{equation}
where
$\zeta=-2\log(-iz+\sqrt{1-z^2})$ and 
$z=(\widetilde{m}i+\sqrt{d})/2$ [see Appendix \ref{app_c}].
The obtained parameters with the NLLS fitting program
are shown in Fig. \ref{fig_Ediffsine}.
In this figure,
the parameters behave continuously except for some discontinuous points
near $\alpha-\alpha_{\rm D}=0.025$ and 0.065.
Comparing Figs. \ref{fig_TT} and \ref{fig_Ediffsine},
we think that the results of the infinite sum version
(Eq. \eqref{eq_infinitesum}) 
as a lattice effect
is more reasonable
than that of the sine wave version (Eq. \eqref{eq_sine}),
although
we have not found a conclusive evidence to support it yet.

\section{Summary and Discussion\label{sec_S}}
In this study,
we have examined the $S=1$ BLBQ model near the AKLT
point.
Analyzing the energy gap of edge states on the basis of the S-A
prescription,\cite{Sorensen-Affleck}
we have shown that
our numerical results support Eq. \eqref{eq_Ediff},
{\it i.e.}
Eq. \eqref{eq_diff}
which is
one of the predictions for the static structure factor
concerned with the C-IC change.
The energy gap of edge states is more manageable than
the correlation function because the singularities are different among
them,
and thus
our results are clearer than the previous one.
We have also obtained the incommensurate wave number,
the amplitude and the correlation length.
These results are consistent with the previous result.
\cite{Schollwock-Jolicoeur-Garel}
Our incommensurate wave number $q_{\rm IC}$
is different from the original incommensurate wave number $k_{\rm IC}$
in Sec. \ref{sec_Intro}.
The difference is caused
of our notation;
the prefactor $(-1)^{N+1}$ of the gap $\Delta E_{\rm ST}$
is left apparently.
Two different wave numbers can be related as 
$k_{\rm IC}=\pi \pm q_{\rm IC}$.

We should mention here that
Eq. \eqref{eq_Ediff} is not only numerically supported,
but also it has a physically favorable feature.
From Eq. \eqref{eq_Ediff}, 
one can see
$\Delta E_{\rm ST}(0)=0$ for $L=1$,
{\it i.e.} $N=0$  which means 
no sub-Hamiltonian case in Eq. \eqref{eq_BLBQHamiltonian}.
Although this property is not necessary since overall $G(q,\kappa)$
consists of singular and regular terms as 
Eqs. \eqref{eq_infinitesumG} and \eqref{eq_sineG},
the property $\Delta E_{\rm ST}(0)=0$ seems quite natural physically.

The amplitude $\widetilde{A}$ has been found to be proportional to
$\sqrt{\alpha - \alpha_{\rm D}}$.
This result implies that $\lambda^2 \propto \alpha -\alpha_{\rm D}$
because of Eq. \eqref{eq_Ediff}.
In Appendix \ref{app_a} and Ref. [\onlinecite{Sorensen-Affleck}],
we have only assumed that the interaction $\lambda$ is some real constant.
However, our results suggest that 
$\lambda$ is some {\it complex} constant.
Thus
we have to modify
the assumption for $\lambda$.
Note that 
$\lambda$ is equal to zero at the disordered point,
corresponding to the VBS picture.

Originally
in the S-A prescription,
the singlet-triplet energy gap $\Delta E_{\rm ST}$ depends on 
the Green function,
which is assumed to have a simple pole 
in the upper half-plane and in the lower half-plane.
However,
our results suggest that two poles should be concealed 
in the upper or lower half-plane.
In general,
one of these poles is far from the real axis,
and therefore
the ordinary field theoretic approach, like the nonlinear $\sigma$
model,\cite{Haldane}
appears to succeed in describing the Haldane phase.
Indeed,
if we explain the whole Haldane phase including the C-IC change,
we must consider the four singular points.
Near the AKLT point,
a four-pole structure
becomes explicit in the Green function,
and then the incommensurability occurs in the
incommensurate Haldane subphase.
A prelude to the incommensurability arises
even in the commensurate region.
We have found that 
positions of poles (singularities) included in the Green function
are represented in terms of $(\widetilde{m}, d)$.

We have left some future tasks;
the effective Lagrangian (maybe two components) and the dispersion curve
for the Green function Eq. \eqref{eq_diff_g} ({\it cf.} those for
Eq. \eqref{eq_prod_g} have been obtained in Ref. \onlinecite{Fath-Suto}),
numerical verification of the static structure factor and
the dynamical structure factor.
Although
we treat only the 1D $S=1$ BLBQ model in this paper,
we have obtained similar results 
about the 1D $S=1/2$ NNN model.\cite{Nomura-Murashima}
However,
we need to modify the discussion about the Green function
since a quasiparticle has a magnon-like behavior in $S=1$ models,
whereas a spinon-like behavior in $S=1/2$ models.

\begin{acknowledgments}
The authors thank I. Affleck for introducing his
 paper\cite{Sorensen-Affleck}
and fruitful discussions.
KN thanks U. Schollw{\"o}ck 
for encouragements 
and A. S{\"u}t{\H o} for giving a prompt 
to consider differences from his approach.
The numerical calculation in this work is based on the program
packages TITPACK version 2, developed by H. Nishimori.
This research is partially supported by a Grant-in-Aid for 
Scientific Research (C),
18540376 (2006),
from the Ministry of Education, 
Science, Sports and Culture of Japan. 
\end{acknowledgments}

\appendix

\section{double-valued function $f(z)$ \label{app_dvf}}
In this appendix \ref{app_dvf},
we examine some properties of $f(z)$
(Eq. \eqref{eq_fz})
introduced in Sec. \ref{subsec_ssf}.
\subsection{Choice of branch cuts and related property}
\begin{figure}[h]
\includegraphics[width=8cm]{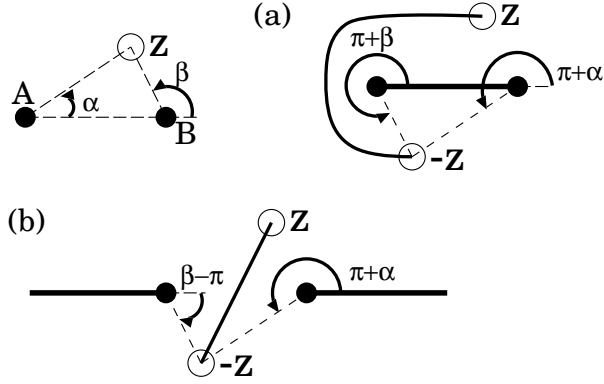}
\caption{Two typical types of branch cut of $f(z)=(z^2 - d)^{-1/2}$.
The  points $A$ and $B$ show $z=-\sqrt{d}$ and $\sqrt{d}$, respectively.}
\label{fig_branch}
\end{figure}

The function
\begin{equation}
f(z)=(z^2-d)^{-1/2}=(z+\sqrt{d})^{-1/2}(z-\sqrt{d})^{-1/2}
\end{equation}
is a double-valued function with two branch points at
$z=-\sqrt{d}$ and $z=\sqrt{d}$.
We can freely choose branch cuts of $f(z)$ 
although the parity of the selected branch cut 
should be compatible with that of $f(z)$.
Typical branch cuts are shown in Fig. \ref{fig_branch}:
(a) both of the branch points are connected, and
(b) each of them are connected to infinite distance.
These different branch cuts bring different parities to $f(z)$.

We consider the case (a) first.
We can carry out the Laurent expansion of $f(z)$ 
around $z=\infty$ when $|z|>\sqrt{|d|}$:
\begin{equation}
(z^2-d)^{-1/2}=\frac{1}{z}\sum_{n=0}^{\infty} \frac{(2n-1)!!}{(2n)!!}
\left(
\frac{d}{z^2}
\right)^{n}.
\end{equation}
It is an odd function with a zero point of order one at infinity.

About the case (b),
we can expand $f(z)$ around $z=0$ when $|z|< \sqrt{|d|}$,
and then we obtain an even function:
\begin{equation}
(z^2-d)^{-1/2}=(-d)^{-1/2}\sum_{n=0}^{\infty} \frac{(2n-1)!!}{(2n)!!}
\left(
\frac{z^2}{d}
\right)^{n}.
\end{equation}

Alternatively,
we can explain their different parities by a graphical way.
Let $z+\sqrt{d}=re^{i \alpha}$ and $z-\sqrt{d}=\rho e^{i \beta}$.
Then 
\begin{equation}
f(z)=r^{-1/2}\rho^{-1/2} e^{-i(\alpha + \beta)/2}.
\end{equation}
In the case (a)
\begin{align}
f(-z)&=r^{-1/2}\rho^{-1/2} e^{-i(\pi + \beta + \pi +
 \alpha)/2}\cr
&= -f(z),
\end{align}
and in the case (b)
\begin{align}
f(-z)&=r^{-1/2}\rho^{-1/2} e^{-i(\beta - \pi + \pi +
 \alpha)/2}\cr
&=f(z).
\end{align}

\subsection{First and second sheets}

\begin{figure}[h]
\includegraphics[width=8cm]{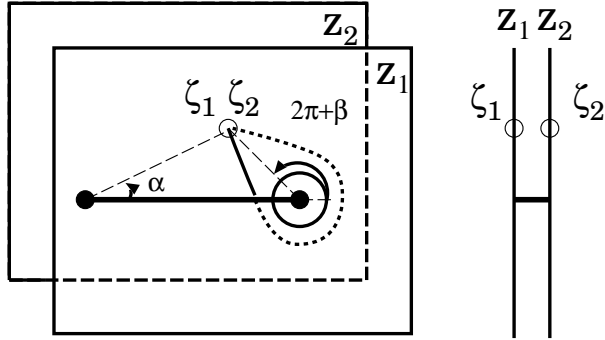}
\caption{Two Riemann sheets of $f(z)=(z^2 - d)^{-1/2}$.
$\zeta_1$ is a point on the first sheet ($z_1$-plane), 
and $\zeta_2$ is on the second sheet ($z_2$-plane).}
\label{fig_riemann}
\end{figure}

The Riemann surface of $f(z)$ consists of two Riemann sheets.
Here, we consider a relation between
the first and second Riemann sheets ($z_1$- and $z_2$-plane, respectively),
although we focus on the case that both of branch points are connected by
a branch cut.
As shown in Fig. \ref{fig_riemann},
let $\zeta_1$ and $\zeta_2$ be a point on the $z_1$- and $z_2$-plane,
respectively, although
these two points have the identical coordinate.
A similar discussion with the previous subsection
can be applied to the case of $\zeta_1 \to \zeta_2$. 
Then we find
\begin{equation}
f(\zeta_2) = -f(\zeta_1).
\end{equation}

\section{Field theoretic approach for edge states\label{app_a}}
In this appendix \ref{app_a} we reproduce the S{\o}rensen and Affleck prescription.
\cite{Sorensen-Affleck}
They start from the nonlinear $\sigma$ (NL$\sigma$)
model.
\cite{Haldane}
Since an effective field model is not clear in our case,
it is not possible to apply this model as it is near the AKLT point.
However,
we can develop a similar discussion
if we assume a Green function 
$G(q,\kappa)$.
The Green function is determined from discussions 
in Secs. \ref{sec_SSF} and \ref{sec_EG}.
It describes some massive free boson fields
$\boldsymbol{\phi}(x,\tau)$:
\begin{equation}
\boldsymbol{\phi}(x,t)=\int \frac{dq}{\sqrt{2\pi}\sqrt{2\omega_q}}
\left(
{\bf a}(q)e^{iqx -i\omega_q t} + {\bf a}^{\dagger}(q)e^{-iqx+i\omega_q t}\right),
\end{equation}
where ${\bf a}(q)$ is a bose operator which satisfies 
$[{\bf a}(q),{\bf a}^{\dagger}(q')]=\delta(q-q')$,
and 
$\omega_q$ is obtained from the Green function.
Vacuum expectation values among two different boson fields 
are calculated as the following:
\begin{equation}
\langle \boldsymbol{\phi}(x,t) \cdot \boldsymbol{\phi}(0,0)\rangle
=
\int \frac{dq}{2\pi}\frac{ e^{iqx-i\omega_q t}}{2\omega_q}.
\end{equation}

\subsection{Static structure factor}
Static structure factor $S(q)$ is defined as the Fourier transform of a equal-time
correlation function:
\begin{equation}
\langle \boldsymbol{\phi}(x,0)\cdot \boldsymbol{\phi}(0,0)\rangle
=
\int \frac{dq}{2\pi} S(q)e^{iqx}.
\end{equation}
Therefore,
we can relate the static structure factor with $\omega_q$:
\begin{equation}
S(q)=\frac{1}{2\omega_q}.
\end{equation}

\subsection{Green function}

Green function is defined as the time-ordered expectation value:
\begin{equation}
iG(x,t)=T\langle \boldsymbol{\phi}(x,t)\cdot\boldsymbol{\phi}(0,0)\rangle.
\end{equation}
Using the Wick rotation
\begin{equation}
t=-i\tau,\quad \omega = -i\kappa, 
\end{equation}
where $\omega t = - \kappa \tau$,
and
the step function
\begin{equation}
\theta(\tau)=\frac{1}{2\pi}\int d\alpha \frac{e^{i\alpha\tau}}{i\alpha},
\end{equation}
we then find
\begin{align}
\int \frac{d\kappa dq}{(2\pi)^2} e^{iqx+i\kappa \tau} iG(q,\kappa)
&= 
\int \frac{d\alpha dq}{2(2\pi)^2} 
\frac{1}
{i\alpha \omega_q}
(e^{iqx+i(-i\omega_q+\alpha) \tau} + e^{-i(-i\omega_q+\alpha) \tau-iqx}),
\nonumber \\
&=\int \frac{d\kappa dq}{(2\pi)^2} e^{iqx+i\kappa \tau}
\frac{i}{\kappa^2 +\omega_q^{2}},
\end{align}
where $\kappa$ is a imaginary frequency. 
The Green function $G(q,\kappa)$ associates with $\omega_q$ as
\begin{equation}
G(q,\kappa)=\frac{1}{\kappa^2 + \omega_q^2}.
\end{equation}

\subsection{Perturbation theory}

Using the Green function $G(q,\kappa)$,
we can describe a free part of the action.
\begin{equation}
S(\boldsymbol{\phi})=\frac{1}{2}\int \frac{d\kappa dq}{(2\pi)^2} G^{-1}(q,\kappa) 
\widetilde{\boldsymbol{\phi}}^2e^{iqx+i\kappa \tau},
\end{equation}
where $\widetilde{\boldsymbol{\phi}}$ is the Fourier transform of
$\boldsymbol{\phi}$.

Next,
we take the edge effects into consideration.
The open boundaries have the effect of leaving a $S=1/2$ 
degree of freedom at each end of chain.
The edge spins will interact with the rest of the system.
To consider this effect,
we assume the following interaction:
\begin{equation}
{\cal H}_{\rm I} = \lambda [\boldsymbol{\phi}(1)\cdot{\bf S'}_1
+(-1)^{L-1} \boldsymbol{\phi}(L)\cdot{\bf S'}_L],
\end{equation}
where $\lambda$ is weak coupling constant. 
${\bf S'}_1$ and ${\bf S'}_L$ are two $S=1/2$ excitations known
to exist at the end of 
the open chain\cite{Kennedy,Hagiwara-Katsumata,Glarum-Geschwind,Mitra}.
The sign in front of the second term comes from
the reason that
we consider the boson field $\boldsymbol{\phi}$ with the wave number $\pi$.

Carrying out the ordinary Gaussian integral,
we can obtain an effective action $S_{\rm eff}({\bf S'}_1, {\bf S'}_L)$.
\begin{equation}
\int {\cal D}\boldsymbol{\phi} 
e^{-S(\boldsymbol{\phi}) + \int d \tau dx 
{\bf J}(x,\tau)\cdot\boldsymbol{\phi}(x,\tau)}
=Ce^{-S_{\rm eff}}, \label{gaussian}
\end{equation}
where ${\bf J}(x,\tau)=\lambda[{\bf S'}_1\delta(x-x_1)
+(-1)^{L-1}{\bf S'}_L\delta(x-x_L)]$.
Then we find
\begin{equation}
S_{\rm eff}=
(-1)^{L}\lambda^2
{\bf S'}_1\cdot{\bf S'}_L 
\int d\tau_1 d\tau_L
\frac{dq d\kappa}{(2\pi)^2}
G(q,\kappa)
e^{iq(L-1) + i\kappa (\tau_L - \tau_1)}.
\end{equation}
The constant $C$ in Eq. \eqref{gaussian} contains the divergent 
{\it self-energy} that comes from terms with both arguments 
included in 
the Green function 
on the same source world-line.
These correspond to virtual $\boldsymbol{\phi}$ particles 
that are emitted and absorbed by the same source.
We are not interested in these,
but only in the variation in the vacuum energy as a function of 
the separation of the sources.

In this appendix,
we have not consider an imaginary time dependency of ${\bf S'}_{1,L}$
since such a dependency has so far been unclear.

\section{Transformation From Green function to Static structure factor\label{app_b}}
In this appendix \ref{app_b},
we will show
that
the static structure factor (Eq. \eqref{eq_diff})
is constructed from the Green
function (Eq. \eqref{eq_diff_g}).

We consider the following integral:
\begin{equation}
\int\frac{e^{i\kappa \tau}d\kappa}{(\kappa - \sqrt{z})(\kappa + \sqrt{z})}
=\frac{i\pi e^{i\tau \sqrt{z}}}{\sqrt{z}},\label{eq_app_1}
\end{equation}
where $\tau > 0$,
$z=re^{i\theta}$, and $\theta = \text{\rm Arg} \, z$ ($0<\theta < 2\pi$).
The right hand side of Eq. \eqref{eq_app_1} 
is a double-valued function
and
has a branch point at $z=0$.

Now we consider $w=\sqrt{z}$.
In general,
$w$ corresponds to $w_1=\sqrt{r}e^{i\theta/2}$
in the upper half $w$-plane
when $0 < \arg z < 2\pi$,
while $w$ corresponds to $w_2=-\sqrt{r}e^{i\theta/2}$
in the lower half $w$-plane
when $2\pi < \arg z < 4\pi$.
Thus Eq. \eqref{eq_app_1} is rewritten as
\begin{equation}
\int\frac{e^{i\kappa \tau}d\kappa}{(\kappa - \sqrt{z})(\kappa+\sqrt{z})}
=
\begin{cases}
i\pi e^{i\tau w_1}/w_1
\quad
&(0 < \arg z <  2\pi),\\
i\pi e^{i\tau w_2}/w_2
\quad
&(2\pi  < \arg z <  4\pi).\label{eq_app_2}
\end{cases}
\end{equation}

Using Eq. \eqref{eq_app_2},
we can show that
\begin{equation}
\lim_{\tau \to 0} \int \frac{d\kappa}{2\pi}
e^{i\kappa \tau} (G_{+}(q,\kappa) + G_{-}(q,\kappa))
=
\frac{Ai}{2\widetilde{m}}(f(q+\widetilde{m}i)-f(q-\widetilde{m}i)).
\end{equation}

\section{Integration of Green function about sine wave
 version\label{app_c}}

Substituting $p(q)=2\sin(q/2)$ for $q$ in Eq. \eqref{eq_diff_g},
we obtain the energy gap behavior of the edge states;
\begin{equation}
\Delta E_{\rm ST}(L-1)=(-1)^{L} \lambda^{2}\frac{A^2}{\widetilde{m}^2}
\int \frac{dq}{2\pi}e^{iq(L-1)}
\{G^{+}(p(q),\kappa) + G^{-}(p(q),\kappa)\}
\label{eq_app_3_1}
\end{equation}
Using the formula $\sin ^{-1} z = i \log (-iz +  \sqrt{1-z^2})$,
we can integrate the right hand side of Eq. \eqref{eq_app_3_1} over $q$.
After the integration,
we find
\begin{equation}
\Delta E_{\rm ST}(L-1)=(-1)^{L}\widetilde{A} e^{\Re(\zeta) (L-1)}
\sin(\Im(\zeta)(L-1)),
\end{equation}
where $\zeta=-2\log(-iz+\sqrt{1-z^2})$ and 
$z=(\widetilde{m}i+\sqrt{d})/2$.

\bibliography{biblio}

\end{document}